\documentclass{trbunofficial}
\usepackage{graphicx}
\graphicspath{ {./images/} }

\usepackage[colorlinks=true,linkcolor=blue,citecolor=blue]{hyperref}

\usepackage{array, makecell}
\usepackage{hanging}
\usepackage{anyfontsize}
\usepackage{float}
\usepackage{amsmath,amssymb}
\usepackage{newtxtext,newtxmath}
\usepackage{pifont}
\usepackage{xcolor}

\makeatletter
\newsavebox\myboxA
\newsavebox\myboxB
\newlength\mylenA

\newcommand*\xoverline[2][0.75]{%
    \sbox{\myboxA}{$\m@th#2$}%
    \setbox\myboxB\null
    \ht\myboxB=\ht\myboxA%
    \dp\myboxB=\dp\myboxA%
    \wd\myboxB=#1\wd\myboxA
    \sbox\myboxB{$\m@th\overline{\copy\myboxB}$}
    \setlength\mylenA{\the\wd\myboxA}
    \addtolength\mylenA{-\the\wd\myboxB}%
    \ifdim\wd\myboxB<\wd\myboxA%
       \rlap{\hskip 0.5\mylenA\usebox\myboxB}{\usebox\myboxA}%
    \else
        \hskip -0.5\mylenA\rlap{\usebox\myboxA}{\hskip 0.5\mylenA\usebox\myboxB}%
    \fi}
\makeatother

\AuthorHeaders{Said, Zajdela, and Stathopoulos}
\title{Accelerating the Adoption of Disruptive Technologies: The Impact of COVID-19 on Intentions to Use Autonomous Vehicles}

\author{%
  \textbf{Maher Said}\\
  Department of Civil and Environmental Engineering\\
  Northwestern University\\
  A308 Technological Institute\\
  2145 Sheridan Road, Evanston, IL, 60208, USA\\
  Email: MaherSaid@u.northwestern.edu\\
  \hfill\break
  \textbf{Emma Zajdela}\\
  Department of Engineering Sciences and Applied Mathematics\\
  Northwestern University\\
  M416 Technological Institute\\
  2145 Sheridan Road, Evanston, IL, 60208, USA\\
  Email: EmmaZajdela@u.northwestern.edu\\
  \hfill\break%
  \textbf{Amanda Stathopoulos}\\
  Department of Civil and Environmental Engineering\\
  Northwestern University\\
  A312 Technological Institute\\
  2145 Sheridan Road, Evanston, IL, 60208, USA\\
  Tel: +1 847-491-5629, Fax: +1 847-491-4011\\
  Email: a-stathopoulos@northwestern.edu\\
}

\TotalWords{8434}

\begin{document}

\maketitle

\section{Abstract}

One of the most notable global transportation trends is the accelerated pace of development in vehicle automation technologies. Uncertainty surrounds the future of automated mobility as there is no clear consensus on potential adoption patterns, ownership versus shared use status and travel impacts. Adding to this uncertainty is the impact of the COVID-19 pandemic that has triggered profound changes in mobility behaviors as well as accelerated adoption of new technologies at an unprecedented rate. Accordingly, this study examines the impact of the COVID-19 pandemic on willingness to adopt the emerging technology of autonomous vehicles (AVs). Using data from a survey disseminated in June 2020 to 700 respondents in the United States, we perform a difference-in-difference regression to analyze the shift in willingness to use autonomous vehicles as part of a shared fleet before and during the pandemic. The results reveal that the COVID-19 pandemic has a positive and highly significant impact on the consideration of using autonomous vehicles. This shift is present regardless of tech-savviness, gender or urban/rural household location. Individuals who are younger, left-leaning and frequent users of shared modes of travel are expected to become more likely to use autonomous vehicles once offered. Understanding  the  effects  of  these  attributes  on  the  increase  in  consideration  of  AVs  is important for policy making, as these effects provide a guide to predicting adoption of autonomous vehicles - once available - and to identify segments of the population likely to be more resistant to adopting AVs.

\vspace{\medskipamount}
\noindent\textbf{Keywords:} autonomous vehicles; self-driving; adoption; technology; COVID pandemic; difference-in-difference

\newpage

\section{Introduction}
Over the last 60 years, researchers have questioned how new technologies are adopted as they are introduced to the market \cite{rogers62}. Some of these technologies have had a major impact not only on everyday lives, but also on society, businesses and the economy. The business opportunities presented by major technological leaps have pushed firms and entrepreneurs to innovate in a variety of fields in order to be at the forefront of the new wave of adopted technologies \cite{lai17}. One such technology that has yet to come to full fruition is the autonomous vehicle (AV). Following the limited success of DARPA’s Grand Challenge  in 2004, Google began developing its own autonomous car project in 2009, today known as Waymo \cite{vance04,waymo18}. However, even with more than 10 million miles of automated driving in real world conditions and billions of miles in simulation, it remains unclear when one of the earlier pioneers in autonomous driving will be market ready \cite{etherington19,hawkins19}. On the other hand, Tesla adopted a more aggressive approach to AV development, announcing Autopilot capability in 2014 and introducing it to the market in 2015 through a software update to its fleet of vehicles, providing partial to conditional automation \cite{white14}. Today, Tesla vehicles have the hardware necessary for complete autonomy \cite{golson16} but still require the software advancements necessary for these capabilities. Other manufacturers have made claims about deploying vehicles with some form of autonomy within the next few years \cite{faggella20}.

With complete autonomy potentially a close reality, notwithstanding predictions that the technology will not be available until decades from now \cite{parkin16, lavasani16, litman17}, understanding adoption of the technology is increasingly important \cite{harb2021we}. It is hypothesized that AVs, like many of the ground-breaking technologies before it, will have significant benefits to society and the economy - as well as potential drawbacks \cite{fagnant15, harper16, harb2021we}. The ease of use of AVs and expected benefits, however, do not necessarily ensure prompt adoption of the technology \cite{bansal16, vallet16}. Given the novelty of this technology, potential safety implications and mistrust in untested technologies, mass-adoption may be slow and take years.

Adding to the uncertainty of future adoption patterns is the ongoing global COVID-19 pandemic, triggering massive changes in behavior. The pandemic has disrupted behavior in almost every facet of society and everyday life, from business, to education, healthcare and retail services \cite{clipper20, greenhow21, belzunegui20}. Most of these disruptions have been in the form of shifting activities from involving in-person interactions to reducing or removing these interactions altogether, such as shifting to telework or e-learning or ordering groceries online instead of visiting the grocery store. On the whole, the pandemic has been an accelerator of technological adoption. This paper hypothesizes that this acceleration in technological adoption is not only limited to current technologies, but could alter sentiment towards and consideration of future technologies, specifically AVs in this study. Additionally, concern of virus transmission could serve as a catalyst for increased preference towards AVs, especially among users of shared modes of transport, by providing a transportation alternative that does not require a driver or any human interaction. An important unanswered question is whether or not the increased AV consideration triggered will remain in the post-COVID era, inviting important research on how future rates of AV adoption are shaped by the pandemic.

Consequently, this study has two goals. The first is to gain understanding of adoption of novel transportation technologies by analyzing the hypothetical willingness to use future automated modes of transport. The second goal is to deepen the understanding of the impact of major life disruptions on technology adoption. Here, this is analyzed by studying the role of the COVID-19 pandemic and related experiences on AV adoption. Specifically, through a difference-in-difference regression experiment, this study estimates the shift in willingness to adopt autonomous vehicle use in the wake of the COVID-19 pandemic. Based on 691 U.S. responses to a 2020 web survey\footnote{The survey was distributed to 700 participants, with 9 responses removed in the process of data cleaning as described later in this paper.}, we identify several factors that affect the shift in adoption willingness. We find that the COVID-19 pandemic has had a significant net positive effect on the willingness to use autonomous vehicles. This effect is more pronounced for younger, politically left-leaning individuals and typical users of shared modes of travel, such as ridehailing and taxi.

The remainder of this paper is organized as follows. The following section presents a brief literature review on technology adoption during COVID-19 and on adoption of autonomous vehicles. Next, the data collection process is described and preliminary insights are provided. The fourth and fifth sections present the modeling results and discussion. Finally, the paper is then concluded with remarks on limitations and future research.

\section{Literature}

In March 2020, the World Health Organization declared the spread of COVID-19 a global pandemic \cite{who20}. Being multiple times more contagious than seasonal influenza and having more severe symptoms and fatality rate \cite{sanche20, earl21}, governments and cities around the world started mandating  shelter-in-place orders as well as social distancing. These orders have had significant effects on lives and communities around the world, heavily restricting mobility \cite{shamshiripour20} and economic activities \cite{ozili20} during the peaks of the pandemic. 

These restrictions as well as wariness towards the virus resulted in rapid shifts in behaviors throughout multiple facets of everyday life, such as work, education, travel and leisure. As a result of these new constraints and changes in consumer demand, technological innovations have emerged and been broadly adopted, including video-conferencing tools for telework, virtual performances in entertainment, grocery delivery services, telehealth and e-learning \cite{brem21,yan20,vargo21,mouratidis21}. For example, the COVID-19 pandemic forced many companies and employees to adapt to a remote work environment and spurred a sharp increase in the use of technologies enabling virtual communication such as Zoom, Microsoft Teams and Slack \cite{leonardi20}. The same is true in education as students and teachers of all ages transitioned to remote classrooms, thus adopting digital tools used to provide a mix of synchronous and asynchronous instruction including emails, video-conferencing and learning management systems \cite{greenhow21}. In the healthcare sector, the pandemic led to the development and adoption of technology solutions providing safety through physical distance for patients and healthcare providers alike, including telehealth visits as well as robotics to perform tasks such as taking a patient's temperature or disinfecting rooms \cite{clipper20,budd20,golinelli20}.

Fearfulness around contracting the virus as well as social distancing guidelines caused a drastic shift in mobility patterns during the pandemic \cite{shamshiripour20}, which brought about the constriction of activities requiring in-person interactions with non-household members and an increase in use of no-contact options. Consequently, consumers became familiar with and willing to adopt no-contact deliveries, curbside pick-up \cite{brem21,charlebois21,unnikrishnan21} as well as autonomous delivery vehicles for groceries, specifically sidewalk robots and mobile parcel lockers, as they were perceived to be safer due to the lack of a driver in the vehicle \cite{kapser21}. Similarly, consumers preferred mobility modes that were perceived as having the lowest risk of exposure, namely a personal vehicle, walking and biking as opposed to modes that were perceived as the riskiest: public transportation, taxi and ride-hailing \cite{shamshiripour20,barbieri21,dingil21}.

A future possible no-contact alternative to the three perceived riskiest modes, particularly for individuals who do not own a personal vehicles and who cannot adopt active mobility modes, is the emerging technology of autonomous vehicles (AVs), especially as part of a shared fleet. The large interest in AVs goes beyond its benefits with regards to curtailing disease transmission and is a result of the disruptive yet potential beneficial changes it is expected to bring to users and to transportation systems. Despite many questions in regards to drawbacks of the technology (such as increased vehicle-miles traveled) as well as barriers to implementation and mass-market penetration \cite{fagnant15}, AVs are posited to have benefits ranging from safety, to fuel-efficiency, congestion mitigation, access to mobility for underserved and excluded groups, and others \cite{fagnant15,harper16}. Nonetheless, the benefits of AVs will not be realized until the technology is adopted at sufficient scale \cite{talebpour16}. 

The obstacles to prompt market penetration are highlighted throughout the literature of adoption of autonomous vehicles, summarized in recent review articles including the ones by Alawadhi et. al. \cite{alawadhi20} and Pigeon, Alauzet and Paire-Ficout \cite{pigeon21}. Studies have shown that trust in the technology is one of the most important factors impacting AV adoption \cite{panagiotopoulos18,hegner19}. In a U.S. survey, 82\% of respondents consider safety as the most important factor affecting their perception and adoption of autonomous vehicles, while only 6\% are more concerned with cost \cite{bansal16}. Other factors include the context in which the autonomous vehicle use would take place, demographics \cite{payre14} and perceived usefulness \cite{hegner19}. 

In summary, a key takeaway from the literature is that the COVID-19 pandemic has precipitated the adoption of many new technologies. Although AVs are not yet fully operational, studies have identified factors that contribute to AV adoption, of which one of the most important is risk perception. It has been hypothesized that COVID-19 could increase trust in AVs \cite{ribeiro21} and thus accelerate willingness to adopt them. This study contributes to the literature by investigating how COVID-19 impacts perception and future adoption of AV services as a no-contact alternative to modes perceived as riskier from a sanitary perspective.

\section{Survey Design \& Data Collection}
The data for this study was collected using a web survey designed on Qualtrics and disseminated through the Prolific platform to 700 contiguous U.S. respondents in early June 2020. Although online sampling can result in self-selection or coverage bias, specifically related to internet access \cite{mercer17}, recent evidence indicates that online samples are more diverse and often comparable in quality to traditional survey samples \cite{casler13, smith15, kees17}.

\subsection{Survey Design}
The main purpose of the survey is to collect information and data on the shift in likelihood of adopting autonomous vehicles as a result of the COVID-19 pandemic. The authors hypothesize that wariness towards in-person interactions is a catalyst for increased consideration of automated modes of travel that minimize the human factor. In order to capture this shift in likelihood, a difference-in-difference experiment is designed. Given the role of in-person interactions - or lack thereof - during the COVID-19 pandemic, the experiment is anchored on shared modes of travel, where at least one unfamiliar person (the driver) is present in the vehicle. Using shared modes as benchmarks, specifically ride-hailing and taxi, the experiment is also more generalizable and relatable to respondents regardless of car-ownership status. Additionally, using shared modes removes deliberations specific to owning an AV and would allow for a more focused assessment of the sentiment towards the technology regardless of the complexities related to ownership, specifically financially. Accordingly, as part of the experiment, respondents are asked the following two questions on a hypothetical ride-hailing AV service,
\begin{enumerate}
  \item "Assume that you need to make a trip for an essential purpose \underline{\textbf{DURING}} the outbreak. How likely are you to request a ride-hailing service \underline{that uses automated cars} and has no driver or other passengers in the vehicle?"
  \item "Assume that you needed to make a trip for an essential purpose \underline{\textbf{BEFORE}} the outbreak. How likely would you have been to request a ride-hailing service \underline{that uses automated cars} and has no driver or other passengers in the vehicle?"
\end{enumerate}
The responses are measured on a 5-point Likert scale ranging from \textit{very unlikely (1)}, to \textit{neither likely nor unlikely (3)}, to \textit{very likely (5)}. A 5-point scale is used in place of a 7-point option to prevent respondent fatigue \cite{rahi17} given that the experiment is a part of a larger survey.

As previously mentioned, using ride-hailing and taxi as control modes, respondents are also asked the following on a similar 5-point Likert scale,
\begin{enumerate}[resume]
  \item "Assume that you need to make a trip for an essential purpose \underline{\textbf{DURING}} the outbreak. How likely are you to request a typical ride-hailing service (such as Uber or Lyft) or taxi service?"
  \item "Assume that you needed to make a trip for an essential purpose \underline{\textbf{BEFORE}} the outbreak. How likely were you to request a typical ride-hailing service (such as Uber or Lyft) or taxi service?"
\end{enumerate}
For the purposes of this experiment, both, ride-hailing and taxi, are suitable control modes \cite{harb18} and are used in conjunction in order to keep the experiment as relatable as possible to most respondents, specifically across different age groups. For the remainder of the paper, these two modes will be jointly referred to as \textit{hailed modes} or \textit{mobility services} for simplicity.

Assuming a counterfactual that the only significant difference between traditional hailed modes and AV ride-hailing is the lack of a human driver, it is possible to measure the effect of COVID-19 on consideration of shared autonomous vehicles while controlling for unobserved mode-specific attributes. It is important to point out that these questions emphasize trips made for essential purposes, as this distinction is likely to influence responses and have an effect on the model presented in later sections of the paper. The survey does not specify what essential purpose trips are, leaving this up to respondents to define according to their lifestyles and state guidelines at the time of the survey.

In addition to the latter experiment, the survey consists of 3 other sections relevant to this study. The first section inquires respondents about COVID-19 factors, covering their status as essential worker, the extent to which the pandemic affected their lives, whether someone in their household has been laid-off, quarantined or hospitalized in this period and the duration that their household has been under stay-at-home order, if at all.

The next section collects information on the respondents’ latent attitudes towards the environment and technology. Respondents are presented with indicator statements such as “I am willing to switch to active modes of transportation (such as walking or cycling) in order to protect the environment” and “Technology is changing society for the better” then asked to rate these statements on a 5-point Likert scale ranging from \textit{strongly disagree (1)}, to \textit{neither agree nor disagree (3)}, to \textit{strongly agree (5)}.

Finally, respondents are presented with questions related to their socioeconomic status and demographics, such as gender, age, ethnicity, employment status, education, household size and income.

\subsection{Sample Description \& Statistics}

For this study, out of 700 responses, 1 response is removed for missing critical data, 1 response is removed for being outside contiguous U.S. (Hawaii) and 7 responses are removed due to being low-quality (straightlining, extreme hastiness, etc.), resulting in 691 usable data points.
Looking at the usable data, responses have been collected from 46 out of 48 states within contiguous U.S. as well as Washington, D.C. The number of responses from the four most represented states and other sample statistics are shown in TABLE \ref{tab:samplestats}. The two unrepresented states, Vermont and Wyoming, are the least populated states, with 0.2\% of the adult population each.

\begin{table}[H]
\captionstyle{\raggedright}
\hangcaption
\caption{Sample Statistics}
\label{tab:samplestats}
\begin{tabular}{lccc}
\Xhline{1.5pt}
\multicolumn{1}{|l|}{Statistics\textsuperscript{†}} &
  \begin{tabular}[c]{@{}c@{}}Sample\\ (responses)\end{tabular} &
  \multicolumn{1}{c|}{\begin{tabular}[c]{@{}c@{}}Sample\\ (\%)\end{tabular}} &
  \multicolumn{1}{c|}{\begin{tabular}[c]{@{}c@{}}U.S. population/\\ other sources$^{*}$ (\%)\end{tabular}} \\ \Xhline{1.5pt}
\multicolumn{1}{|l}{\textbf{State}} & & \multicolumn{1}{p{60pt}|}{} & \multicolumn{1}{c|}{} \\ \hline
\multicolumn{1}{|l|}{\hspace{10 pt}California}                                   & 110 & \multicolumn{1}{c|}{15.9\%} & \multicolumn{1}{c|}{12.0\%} \\
\multicolumn{1}{|l|}{\hspace{10 pt}Texas}                                        & 55  & \multicolumn{1}{c|}{7.9\%}  & \multicolumn{1}{c|}{8.3\%}  \\
\multicolumn{1}{|l|}{\hspace{10 pt}New York}                                     & 52  & \multicolumn{1}{c|}{7.5\%}  & \multicolumn{1}{c|}{6.2\%}  \\
\multicolumn{1}{|l|}{\hspace{10 pt}Florida}                                      & 46  & \multicolumn{1}{c|}{6.6\%}  & \multicolumn{1}{c|}{6.7\%}  \\ \hline
\multicolumn{4}{|l|}{\textbf{Gender}}                                                                                                           \\ \hline
\multicolumn{1}{|l|}{\hspace{10 pt}Male}                                         & 346 & \multicolumn{1}{c|}{50.1\%} & \multicolumn{1}{c|}{48.7\%} \\
\multicolumn{1}{|l|}{\hspace{10 pt}Female}                                       & 331 & \multicolumn{1}{c|}{47.9\%} & \multicolumn{1}{c|}{51.3\%} \\
\multicolumn{1}{|l|}{\hspace{10 pt}Non-Binary}                                   & 14  & \multicolumn{1}{c|}{2.0\%}  & \multicolumn{1}{c|}{-}       \\ \hline
\multicolumn{4}{|l|}{\textbf{Age}}                                                                                                              \\ \hline
\multicolumn{1}{|l|}{\hspace{10 pt}18-24 years}                                  & 207 & \multicolumn{1}{c|}{30.0\%} & \multicolumn{1}{c|}{12.2\%}  \\
\multicolumn{1}{|l|}{\hspace{10 pt}25-34 years}                                  & 229 & \multicolumn{1}{c|}{33.1\%} & \multicolumn{1}{c|}{17.9\%} \\
\multicolumn{1}{|l|}{\hspace{10 pt}35-44 years}                                  & 136 & \multicolumn{1}{c|}{19.7\%} & \multicolumn{1}{c|}{16.3\%} \\
\multicolumn{1}{|l|}{\hspace{10 pt}45-54 years}                                  & 56  & \multicolumn{1}{c|}{8.1\%}  & \multicolumn{1}{c|}{16.7\%} \\
\multicolumn{1}{|l|}{\hspace{10 pt}55-64 years}                                  & 39  & \multicolumn{1}{c|}{5.6\%}  & \multicolumn{1}{c|}{16.6\%} \\
\multicolumn{1}{|l|}{\hspace{10 pt}65 years or older}                            & 24   & \multicolumn{1}{c|}{3.5\%}  & \multicolumn{1}{c|}{20.2\%}  \\ \hline
\multicolumn{4}{|l|}{\textbf{Political Leaning}}                                                                                                \\ \hline
\multicolumn{1}{|l|}{\hspace{10 pt}Democrat}                                     & 406 & \multicolumn{1}{c|}{60.0\%} & \multicolumn{1}{c|}{51.1\%} \\
\multicolumn{1}{|l|}{\hspace{10 pt}Republican}                                   & 98  & \multicolumn{1}{c|}{14.5\%} & \multicolumn{1}{c|}{41.5\%} \\
\multicolumn{1}{|l|}{\hspace{10 pt}Independent/no preference}                    & 155 & \multicolumn{1}{c|}{22.9\%} & \multicolumn{1}{c|}{7.4\%}  \\
\multicolumn{1}{|l|}{\hspace{10 pt}Other (progressive, liberal, libertarian, …)} & 18  & \multicolumn{1}{c|}{2.6\%}  & \multicolumn{1}{c|}{-}       \\ \hline
\multicolumn{4}{|l|}{\textbf{Income}}                                                                                                           \\ \hline
\multicolumn{1}{|l|}{\hspace{10 pt}\fontsize{10pt}{10pt}$<$ \$25,000}                           & 112 & \multicolumn{1}{c|}{16.6\%} & \multicolumn{1}{c|}{14.4\%} \\
\multicolumn{1}{|l|}{\hspace{10 pt}\$25,000 - \$49,999}                          & 156 & \multicolumn{1}{c|}{23.1\%} & \multicolumn{1}{c|}{19.6\%} \\
\multicolumn{1}{|l|}{\hspace{10 pt}\$50,000 - \$99,999}                          & 254 & \multicolumn{1}{c|}{37.7\%} & \multicolumn{1}{c|}{32.0\%} \\
\multicolumn{1}{|l|}{\hspace{10 pt}\$100,000 - \$149,999}                        & 97  & \multicolumn{1}{c|}{14.4\%} & \multicolumn{1}{c|}{17.3\%} \\
\multicolumn{1}{|l|}{\hspace{10 pt}\fontsize{10pt}{10pt}$\ge$ \$150,000}                                  & 55  & \multicolumn{1}{c|}{8.2\%}  & \multicolumn{1}{c|}{16.7\%} \\ \Xhline{1.5pt}
\multicolumn{4}{l}{\begin{tabular}[c]{@{}l@{}}{\scriptsize \textsuperscript{†} Non-responses per category: state = 0, gender = 14, age = 0, political leaning = 14, income = 17}\\ {\scriptsize $^{*}$ Sources: U.S. Census \cite{census19}: state, gender, age and income}\\ {\scriptsize\hspace{36.2pt}Pew Research 2018 Sample \cite{pew19}: political leaning}\end{tabular}}
\end{tabular}
\end{table}

\pagebreak

Looking at the age distribution, most of the sample (82.8\%) is between the ages of 18 and 45, with mean and median age of 33.4 and 30 years respectively, resulting in a sample leaning towards younger adults. Politically, the sample is Democrat leaning. Excluding respondents who preferred not to answer and other responses, 61.6\% of the sample lean Democrat, 14.9\% lean Republican and 23.5\% consider their political views as independent or have no preference. While data has been collected in a slightly different manner, according to samples surveyed by Pew Research in 2018 \cite{pew19}, the latter percentages are 51.1\%, 41.5\% and 7.4\% respectively. This bias towards Left-leaning partisanship is not, however, attributed to the higher response rates from states that tend to vote Democrat in recent elections, an observation confirmed by simulating elections using the sample distribution and election data \cite{wasserman20}. Instead, this bias is likely the result of self-selection in non-random online surveys \cite{heen14, zhang20, huff15}. Finally, the average and median income for the sample are \$72,100 and \$62,500 annually, compared to \$88,600 and \$62,800 in the 2019 5-year American Community Survey \cite{census20b}. Accounting for the household size, the average and median annual household income for the sample are approximately \$29,300 and \$21,900 annually per household-member, respectively. Additional sample statistcs are provided in TABLE \ref{tab:otherstats}.

\begin{table}
\captionstyle{\raggedright}
\hangcaption
\caption{Additional Sample Statistics}
\label{tab:otherstats}
\begin{tabular}{lccc}
\Xhline{1.5pt}
\multicolumn{1}{|p{218pt}|}{\rule{0pt}{16pt}Statistics} &
  Responses &
  \multicolumn{1}{c|}{Percentage} \\[5pt]\Xhline{1.5pt}
\multicolumn{1}{|l}{\textbf{Occupation}} 
& \multicolumn{1}{p{102pt}}{} & \multicolumn{1}{p{114pt}|}{} \\ \hline
\multicolumn{1}{|l|}{\hspace{10 pt}Employed}                                   & 379 & \multicolumn{1}{c|}{54.8\%} \\
\multicolumn{1}{|l|}{\hspace{30 pt}\textit{Essential worker}}                                   & \textit{121} & \multicolumn{1}{c|}{-} \\
\multicolumn{1}{|l|}{\hspace{30 pt}\textit{Non-essential worker}}                                   & \textit{258} & \multicolumn{1}{c|}{-} \\
\multicolumn{1}{|l|}{\hspace{10 pt}Student}                                        & 126  & \multicolumn{1}{c|}{18.2\%} \\
\multicolumn{1}{|l|}{\hspace{10 pt}Other \textit{(retired, unemployed, etc.)}}                                     & 186  & \multicolumn{1}{c|}{26.9\%}  \\ \hline
\multicolumn{3}{|l|}{\textbf{Highest level of education}}                                                                                                           \\ \hline
\multicolumn{1}{|l|}{\hspace{10 pt}Middle school or no education}                                & 3  & \multicolumn{1}{c|}{0.04\%}  \\
\multicolumn{1}{|l|}{\hspace{10 pt}High school}                                & 175  & \multicolumn{1}{c|}{25.3\%}  \\
\multicolumn{1}{|l|}{\hspace{10 pt}Undergraduate degree or trade school}                                       & 398 & \multicolumn{1}{c|}{57.6\%}  \\
\multicolumn{1}{|l|}{\hspace{10 pt}Graduate degree}                                         & 115 & \multicolumn{1}{c|}{16.6\%} \\ \hline
\multicolumn{3}{|l|}{\textbf{Household type}}                                                                                                              \\ \hline
\multicolumn{1}{|l|}{\hspace{10 pt}Urban}                                  & 197 & \multicolumn{1}{c|}{28.5\%}   \\
\multicolumn{1}{|l|}{\hspace{10 pt}Suburban}                                  & 402 & \multicolumn{1}{c|}{58.2\%} \\
\multicolumn{1}{|l|}{\hspace{10 pt}Rural} & 92 & \multicolumn{1}{c|}{13.3\%}  \\ \hline
\multicolumn{3}{|l|}{\textbf{Race/Ethnicity}}                                                                                                              \\ \hline                                 \multicolumn{1}{|l|}{\hspace{10 pt}White}                                     & 466 & \multicolumn{1}{c|}{67.4\%} \\
\multicolumn{1}{|l|}{\hspace{10 pt}Asian}                                   & 111  & \multicolumn{1}{c|}{16.1\%}\\ 
\multicolumn{1}{|l|}{\hspace{10 pt}Black}                                   & 56  & \multicolumn{1}{c|}{8.1\%}\\
\multicolumn{1}{|l|}{\hspace{10 pt}Hispanic or Latino}                                   & 46  & \multicolumn{1}{c|}{6.7\%}\\
\multicolumn{1}{|l|}{\hspace{10 pt}Other}                                  & 12  & \multicolumn{1}{c|}{1.7\%}\\ \hline
\multicolumn{3}{|l|}{\textbf{Typical modes of travel}\textsuperscript{‡}}                                                                                                              \\ \hline                                 \multicolumn{1}{|l|}{\hspace{10 pt}Private vehicle \textit{(incl. motorcycle)}}                                     & 596 & \multicolumn{1}{c|}{86.3\%} \\
\multicolumn{1}{|l|}{\hspace{10 pt}Public transportation}                                   & 157  & \multicolumn{1}{c|}{22.7\%}\\ 
\multicolumn{1}{|l|}{\hspace{10 pt}Hailed rides \textit{(ridehailing or taxi)}}                                   & 123  & \multicolumn{1}{c|}{17.8\%}\\
\multicolumn{1}{|l|}{\hspace{10 pt}Active travel \textit{(walking or cycling)}}                                  & 287  & \multicolumn{1}{c|}{41.5\%}\\ \hline
\multicolumn{3}{|l|}{\textbf{Impact of COVID}}  \\ \hline
\multicolumn{1}{|l|}{\hspace{10 pt}\makecell[l]{One or more household member \\ quarantined during COVID}}                                     & 101 & \multicolumn{1}{c|}{14.6\%} \\
\Xhline{1.5pt}
\multicolumn{3}{l}{\scriptsize \textsuperscript{‡} Respondents are allowed to select multiple modes; percent values sum up to over 100\%.}
\end{tabular}
\end{table}

\begin{figure}[H]
\includegraphics[scale=1.02]{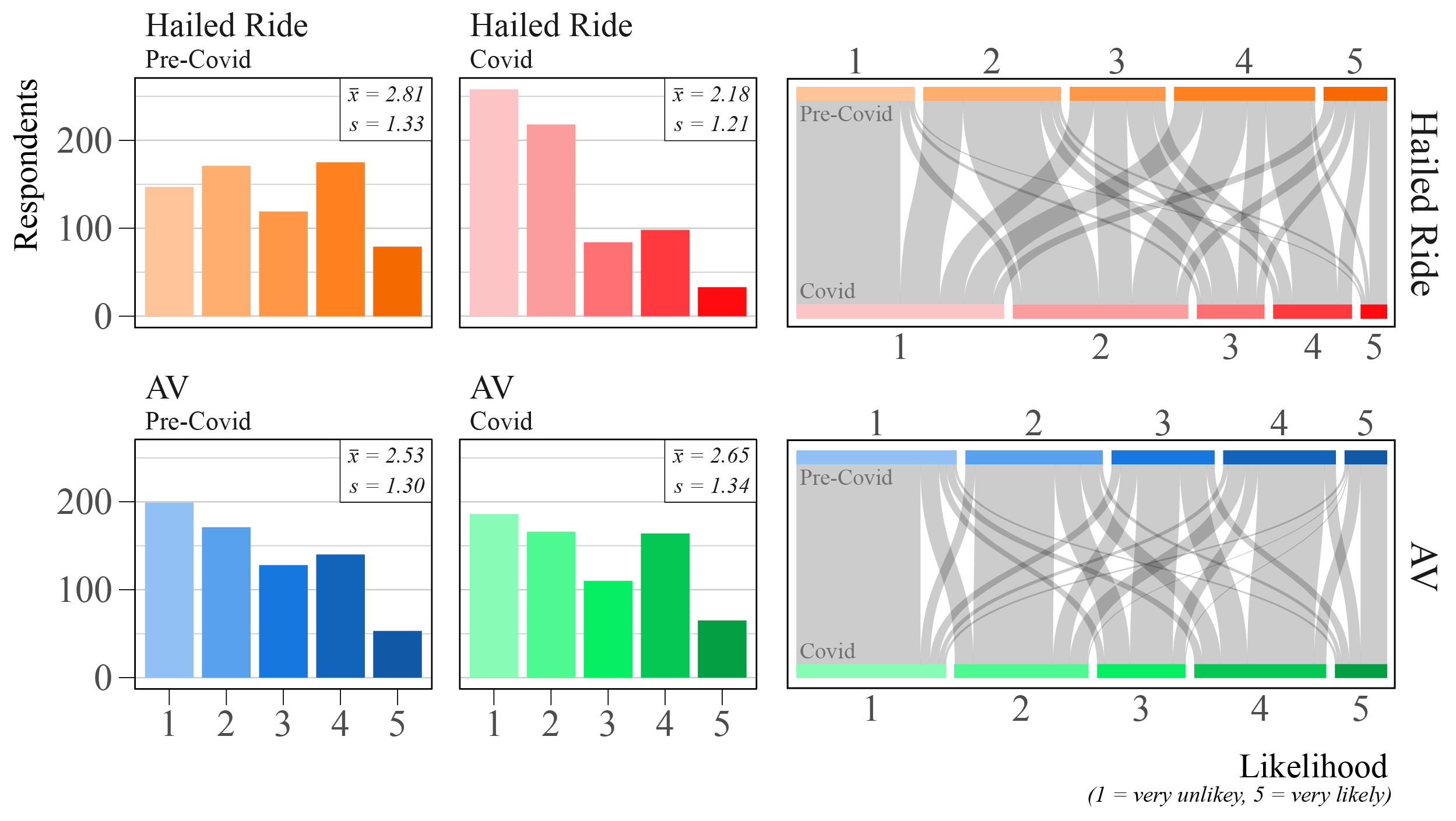}
\captionstyle{\raggedright}
\hangcaption
\caption{(a) Response distributions and (b) flow diagrams for likelihood of using hailed ride and AV ride-sharing before and during the COVID-19 pandemic}
\label{fig:llhist}
\centering
\end{figure}

As for the response patterns of the likelihood of using typical hailed services before and during the COVID pandemic, the average response score shifts from 2.81 (closest to the scale midpoint of \textit{neither likely nor unlikely}) to 2.18 (closest to \textit{unlikely}) respectively. In other words, on average, a decrease in the stated likelihood to use hailed modes is observed during the pandemic (significant with p = 0.000). As shown in Figure \ref{fig:llhist}, 42.1\% of respondents claim a decrease in likelihood of using hailed ride due to COVID. The largest share of respondents (48.5\%) remain at pre-COVID levels.

For the hypothetical AV ride-sharing service, the average response score increases from 2.53 (between \textit{unlikely} and \textit{neither likely or unlikely}) to 2.65 (closer to \textit{neither likely or unlikely}). Nonetheless, this difference is not statistically significant, with a Mann-Whitney-Wilcoxon Test p-value of 0.125.
This can be observed visually in Figure \ref{fig:llhist} through the small shifts in likelihood of using the hypothetical AV service as a result of COVID. Indeed, 61.1\% of respondents believe that their likelihood of using this service would remain unaffected by COVID-19. Yet, 23.4\% claim an increase in likelihood of using AV ride-sharing as a result of the pandemic, versus 15.5\% claiming a decrease in likelihood.

Statistics for the attitudinal indicators are provided in Table \ref{tab:indicstats}. Although some of the indicators for attitudes towards the environment revolve around package delivery (relevant to another study using other sections of the survey data), these indicators are still related to the general construct of \textit{environmental consciousness} and, therefore, considered in this study.

\begin{table}[H]
\captionstyle{\raggedright}
\hangcaption
\caption{Mean and Standard Deviation for Attitudinal Construct Indicators\vspace{3pt}}
\label{tab:indicstats}
\begin{tabular}{lcc}
\Xhline{1.5pt}
\multicolumn{1}{|l|}{Indicator\textsuperscript{†}} &
  Mean &
  \multicolumn{1}{c|}{\makecell{Standard\\deviation}} \\
\Xhline{1.5pt}
\multicolumn{1}{|l}{\textbf{Attitude towards technology}} & \multicolumn{1}{p{70pt}}{} & \multicolumn{1}{p{70pt}|}{} \\ \hline
\multicolumn{1}{|l|}{\hspace{10 pt}Technology is changing society for the better.}                                         & 3.89 & \multicolumn{1}{c|}{0.78} \\
\multicolumn{1}{|l|}{\hspace{10 pt}I am excited to learn about new technologies in the market.}                                       & 4.16 & \multicolumn{1}{c|}{0.82}  \\
\multicolumn{1}{|l|}{\hspace{10 pt}I pay more to get more technologically advanced products.}                                & 3.33  & \multicolumn{1}{c|}{1.06}  \\
\multicolumn{1}{|l|}{\hspace{10 pt}I use the internet daily for chatting and entertainment.}                                & 4.56  & \multicolumn{1}{c|}{0.70}  \\ \hline
\multicolumn{3}{|l|}{\textbf{Attitude towards the environment}}                                                                                                           \\ \hline
\multicolumn{1}{|l|}{\hspace{10 pt}\makecell[l]{I am willing to switch to active mode of transportation \\ \hspace{10 pt}(such as walking or cycling) in order to protect \\ \hspace{10 pt}the environment.}}                                        & \makecell{3.44\\\phantom{}\\\phantom{}} & \multicolumn{1}{c|}{\makecell{1.06\\\phantom{}\\\phantom{}}} \\
\multicolumn{1}{|l|}{\hspace{10 pt}\makecell[l]{I would select more environmentally friendly package \\ \hspace{10 pt}delivery options at the cost of slower delivery.}}                                       & \makecell{3.57\\\phantom{}} & \multicolumn{1}{c|}{\makecell{1.07\\\phantom{}}}  \\
\multicolumn{1}{|l|}{\hspace{10 pt}\makecell[l]{I prefer to order items online in bulk to minimize the total \\ \hspace{10 pt}number of delivery trips made to my address.}}                                & \makecell{3.77\\\phantom{}}  & \multicolumn{1}{c|}{\makecell{1.05\\\phantom{}}}  \\
\multicolumn{1}{|l|}{\hspace{10 pt}I am concerned with the news about climate change.}                                & 4.08  & \multicolumn{1}{c|}{1.07}  \\
\Xhline{1.5pt}
\multicolumn{3}{l}{\begin{tabular}[c]{@{}l@{}}{\scriptsize \textsuperscript{†} 5-point Likert scale: (1) strongly disagree; (2) disagree; (3) neither agree nor disagree; (4) agree; (5) strongly agree}\end{tabular}}
\end{tabular}
\end{table}

\section{Effect of COVID-19 on Willingness to Use Autonomous Vehicles}

Difference-in-difference regression allows for capturing the impact of a treatment or intervention on a target group with respect to a control group through systematic comparison of observed outcomes. The goal of a difference-in-difference regression is to estimate the isolated treatment effect of a natural non-experimental intervention while eliminating unobserved biases that may be present between groups. Without other covariates, a generic difference-in-difference regression is formulated as,

\begin{equation}
y_{imt} = \beta_{\:0} + \beta_{\:1} \: Target_{im} + \beta_{\:2} \: Treated_{it} + \beta_{\:3} \: (Target_{im} \times Treated_{it}) + \epsilon_{imt}
\end{equation}

\bigskip

\noindent where $y_{imt}$ is an observed outcome for individual $i$ in group $m$ given treatment $t$, $Target_{im}$ is a dummy variable equal to 1 if the outcome corresponds to the target group and 0 otherwise, $Treated_{it}$ is a dummy variable equal to 1 if the outcome corresponds to an observation that has received the treatment and 0 otherwise, $(Target_{im} \times Treated_{it})$ is a binary term equal to the product of the latter two dummy variables, and $\epsilon_{imt}$ is the error term. Here, the \textit{treatment effect} is calculated via the estimated parameter $\beta_{\:3}$.

In the same vein, a difference-in-difference regression model is estimated to calculate the shift in consideration of \textit{AV services} (target group) as a result of the \textit{COVID-19 pandemic} (treatment) while controlling for unobserved mode-specific attributes. Assuming the only key difference between traditional hailed modes of travel and AV ride-sharing services is the lack of a driver, \textit{hailed services} (ridehailing and taxi) are used as a control group in this model. This assumption entails that \textit{Parallel Trends} holds true (i.e. in the absence of treatment, the difference in likelihood of using traditional hailed services and shared AV services remains constant across time). Accordingly, the difference-in-difference model is initially formulated as follows,

\begin{equation}
\begin{split}
LL_{imt} = \beta_{\:0} & + \beta_{\:1} \: (Mode = AV)_{im} + \beta_{\:2} \: (Time = COVID)_{it} \\ & + \beta_{\:3} \: ((Mode = AV)_{im} \times (Time = COVID)_{it}) + \epsilon_{imt}
\end{split}
\end{equation}

\bigskip

\noindent where $LL_{imt}$ is the observed likelihood of individual $i$ using mode $m$ at time $t$, $(Mode = AV)_{im}$ is a dummy variable equal to 1 if the mode is \textit{AV} and 0 otherwise, $(Time = COVID)_{it}$ is a dummy variable equal to 1 if time is \textit{during the COVID pandemic} and 0 otherwise, and $\epsilon_{imt}$ is the error term.

Including covariates such as gender, age and income, the formulation is extended as,

\begin{equation}
\begin{split}
LL_{imt} = \beta_{\:0} & + \beta_{\:1} \: (Mode = AV)_{im} + \beta_{\:2} \: (Time = COVID)_{it} \\ & + \beta_{\:3} \: ((Mode = AV)_{im} \times (Time = COVID)_{it}) + \sum^{K} \beta_k \: X_{ki} + \epsilon_{imt}
\end{split}
\end{equation}

\bigskip

\noindent where each $\beta_k$ denotes a coefficient for socio-economic and choice attributes and $X_{ki}$ is a set of respondent-specific covariates.

For a more detailed treatment of difference-in-difference models, the reader is referred to \citep{wing18}.

Given that ${LL}_{it}$ is an ordinal measure, the difference-in-difference regression is estimated using an ordered logit regression model to appropriately account for the nature of the dependent variable. For information on ordered regression modeling, the reader is referred to \citep{greene10}. The results are presented in TABLE \ref{tab:dndresults} using robust standard errors clustered at the respondent level.

\subsection{Impact of COVID on Consideration of Autonomous Vehicles}
The difference-in-difference coefficients are presented in TABLE \ref{tab:dndresults}. Looking at the coefficient for the mode (\textit{mode = AV}), the model indicates that there is a general propensity among respondents away from autonomous vehicles in comparison to traditional vehicles prior to the COVID pandemic (p = 0.000). The magnitude of the latter coefficient ($\beta$ = -0.441) is roughly half-a-level on the 5-point Likert scale used in this study.
Similarly, the time coefficient (\textit{time = COVID}) indicates a decrease in consideration of hailed ride services due to COVID of roughly a full consideration level ($\beta$ = -0.978, p = 0.000). 

\begin{table}[H]
\captionstyle{\raggedright}
\hangcaption
\caption{Ordered Difference-in-Difference Regression for Shift in Likelihood of AV Use due to COVID-19}
\label{tab:dndresults}
\begin{tabular}{p{0.6\linewidth}ccc}
\Xhline{1.5pt}
\multicolumn{4}{|l|}{\textbf{\textit{Model Statistics:}} \hspace{85 pt} \textit{Number of Observations: 691} \hspace{41 pt} \textit{$\rho^2$: 0.067}}                                                                     \\
\multicolumn{4}{|l|}{\hspace{171 pt} \textit{Log-Likelihood at zero: -4,245.17} \hspace{22 pt} \textit{AIC: 7,954.52}}                                                                                                                  \\
\multicolumn{4}{|l|}{\hspace{171 pt} \textit{Final Log-Likelihood: -3,960.26} \hspace{28 pt} \textit{BIC: 8,055.23}}                                                                                                             \\ \Xhline{1.5pt}
\multicolumn{1}{|l|}{Parameter}                                                            & Coefficient & t-value & \multicolumn{1}{c|}{p-value} \\ \Xhline{1.5pt}
\multicolumn{4}{|l|}{\textbf{Difference-in-Difference Coefficients}}                                                                                       \\ \hline
\multicolumn{1}{|>{\raggedright}p{0.63\linewidth}|}{\hangindent=25pt\hspace{10 pt}\underline{Mode = AV}: Mode is on-demand AV}                                      & -0.441$^{***}$      & -6.24   & \multicolumn{1}{c|}{0.000}   \\
\multicolumn{1}{|>{\raggedright}p{0.63\linewidth}|}{\hangindent=25pt\hspace{10 pt}\underline{Time = COVID}: Time is during COVID pandemic}                          & -0.978$^{***}$      & -13.01  & \multicolumn{1}{c|}{0.000}   \\
\multicolumn{1}{|>{\raggedright}p{0.63\linewidth}|}{\hangindent=25pt\hspace{10 pt}\underline{Treatment Effect}: Effect of COVID on on-demand AV adoption}           & 1.149$^{***}$       & 13.69   & \multicolumn{1}{c|}{0.000}   \\ \hline
\multicolumn{4}{|l|}{\textbf{Socioeconomic and Demographics}}                                                                                              \\ \hline
\multicolumn{1}{|>{\raggedright}p{0.63\linewidth}|}{\hangindent=25pt\hspace{10 pt}\underline{Male}: Gender is male (vs. female and non-binary)}                     & 0.187$^{*}$       & 1.71    & \multicolumn{1}{c|}{0.087}   \\
\multicolumn{1}{|>{\raggedright}p{0.63\linewidth}|}{\hangindent=25pt\hspace{10 pt}\underline{Urban}: Urban household (vs. suburban and rural)}                      & 0.212$^{*}$       & 1.72    & \multicolumn{1}{c|}{0.086}   \\
\multicolumn{1}{|>{\raggedright}p{0.63\linewidth}|}{\hangindent=25pt\hspace{10 pt}\underline{(Age - 45) $\times\:\:$(Age \fontsize{10pt}{10pt}$\ge$ 45)}: Age of respondent (in years); linear coefficient ONLY for respondents 45 years old or older} &
  -0.047$^{***}$ &
  -3.96 &
  \multicolumn{1}{c|}{0.000} \\
\multicolumn{1}{|>{\raggedright}p{0.63\linewidth}|}{\hangindent=25pt\hspace{10 pt}\underline{AsianPacific}: Ethnicity is Asian or Pacific Islander (vs. other ethnicities)} &
  0.343$^{***}$ &
  2.76 &
  \multicolumn{1}{c|}{0.006} \\ \hline
\multicolumn{4}{|l|}{\textbf{Typical Modes of Travel}}                                                                                                     \\ \hline
\multicolumn{1}{|>{\raggedright}p{0.63\linewidth}|}{\hangindent=25pt\hspace{10 pt}\underline{PrivateMode}: Typical mode of travel is private car or motorcycle (vs. transit, walking, cycling and transit)} &
  -0.457$^{***}$ &
  -2.98 &
  \multicolumn{1}{c|}{0.003} \\
\multicolumn{1}{|>{\raggedright}p{0.63\linewidth}|}{\hangindent=25pt\hspace{10 pt}\underline{HailedModes}: Typical mode of travel is taxi or ridesharing (vs. transit, walking, cycling and transit)} &
  1.073$^{***}$ &
  8.26 &
  \multicolumn{1}{c|}{0.000} \\ \hline
\multicolumn{4}{|l|}{\textbf{Political Views}}                                                                                                     \\ \hline
\multicolumn{1}{|>{\raggedright}p{0.63\linewidth}|}{\hangindent=25pt\hspace{10 pt}\underline{DemLeft}: Political view is either leaning Democrat, Democrat or other Left view (vs. independent, leaning Republican or other Right view)} &
  -0.397$^{***}$ &
  -3.32 &
  \multicolumn{1}{c|}{0.001} \\
\multicolumn{1}{|>{\raggedright}p{0.63\linewidth}|}{\hangindent=25pt\hspace{10 pt}\underline{PolViewMissing}: Political view is not reported (vs. reported)}        & -0.115      & -0.263  & \multicolumn{1}{c|}{0.793}   \\ \hline
\multicolumn{4}{|l|}{\textbf{Latent Variable}}                                                                                                             \\ \hline
\multicolumn{1}{|>{\raggedright}p{0.63\linewidth}|}{\hangindent=25pt\hspace{10 pt}\underline{Tech-Savviness}: Attitude towards and comfort with overall technology} & 0.413$^{***}$       & 5.93    & \multicolumn{1}{c|}{0.000}   \\ \hline
\multicolumn{4}{|l|}{\textbf{Impact of COVID}}                                                                                                             \\ \hline
\multicolumn{1}{|>{\raggedright}p{0.63\linewidth}|}{\hangindent=25pt\hspace{10 pt}\underline{Quarantined}: One or more household members have been quarantined because of COVID-19} &
  0.277$^{**}$ &
  2.12 &
  \multicolumn{1}{c|}{0.034} \\ \hline
\multicolumn{4}{|l|}{\textbf{Ordinal Thresholds}}                                                                                                          \\ \hline
\multicolumn{1}{|>{\raggedright}p{0.63\linewidth}|}{\hspace{10 pt}$\mu_{(1|2)}$}                                                                  & -1.798$^{***}$      & -12.21  & \multicolumn{1}{c|}{0.000}   \\
\multicolumn{1}{|>{\raggedright}p{0.63\linewidth}|}{\hspace{10 pt}$\mu_{(2|3)}$}                                                                  & -0.523$^{***}$      & -3.63   & \multicolumn{1}{c|}{0.000}   \\
\multicolumn{1}{|>{\raggedright}p{0.63\linewidth}|}{\hspace{10 pt}$\mu_{(3|4)}$}                                                                  & 0.281$^{*}$       & 1.95    & \multicolumn{1}{c|}{0.051}   \\
\multicolumn{1}{|>{\raggedright}p{0.63\linewidth}|}{\hspace{10 pt}$\mu_{(4|5)}$}                                                                  & 1.985$^{***}$       & 13.01   & \multicolumn{1}{c|}{0.000}   \\ \Xhline{1.5pt}
\multicolumn{4}{l}{\scriptsize $^{***}$ = significant at 0.01 level; $^{**}$ = significant at 0.05 level; $^{*}$ = significant at 0.10 level}                                                                                                                             
\end{tabular}
\end{table}

\pagebreak

The difference-in-difference treatment effect (\textit{treatment effect}), however, has a positive and highly significant coefficient (p = 0.000). This indicates that the COVID-19 pandemic has had a positive and significant impact on the consideration of autonomous vehicles. Specifically, the magnitude of this impact ($\beta$ = +1.149) is slightly over a full level on the likelihood scale used in this study. In other words, the impact of the pandemic on AV consideration is roughly the same as transitioning from being \textit{neither likely nor unlikely} to use an AV for making an essential trip to being \textit{likely} to do so. As discussed in the previous section, visually, this positive treatment effect can be observed in Figure \ref{fig:llhist}, where hailed ride consideration is lowered due to the pandemic, yet the consideration of the hypothetical on-demand AV service remains relatively unaffected on average.

\subsection{Control Covariates}
As shown in TABLE \ref{tab:dndresults}, the difference-in-difference model is estimated along a number of other covariates. Although difference-in-difference models control for unobserved mode-specific attributes, the selected covariates control for user-specific attributes, specifically: \textit{gender}, \textit{place of residence}, \textit{age}, \textit{ethnicity}, \textit{typical mode of travel}, \textit{political views}, \textit{tech-savviness} and \textit{major impacts of COVID on the respondent}. The interpretation of the respective coefficients for these covariates, nonetheless, is ambiguous. Given that the dependent variable ${LL}_{it}$ is a measure of likelihood irrespective of mode or time, the latter coefficients measure the effect of a covariate on consideration of both, hailed and on-demand AV, for both, pre-COVID and during COVID. As a simplification, these coefficients can be considered as a general measure of the effect of these covariates on acceptability of on-demand mobility services, without accounting for autonomy or the pandemic. Given the built-in ambiguity in interpretation, we focus our discussion on a few select coefficients below, followed by a deeper analysis assessing the treatment effects on AV consideration by user-specific attributes.

\subsubsection{Age}
\begin{figure}[H]
\includegraphics[scale=.9]{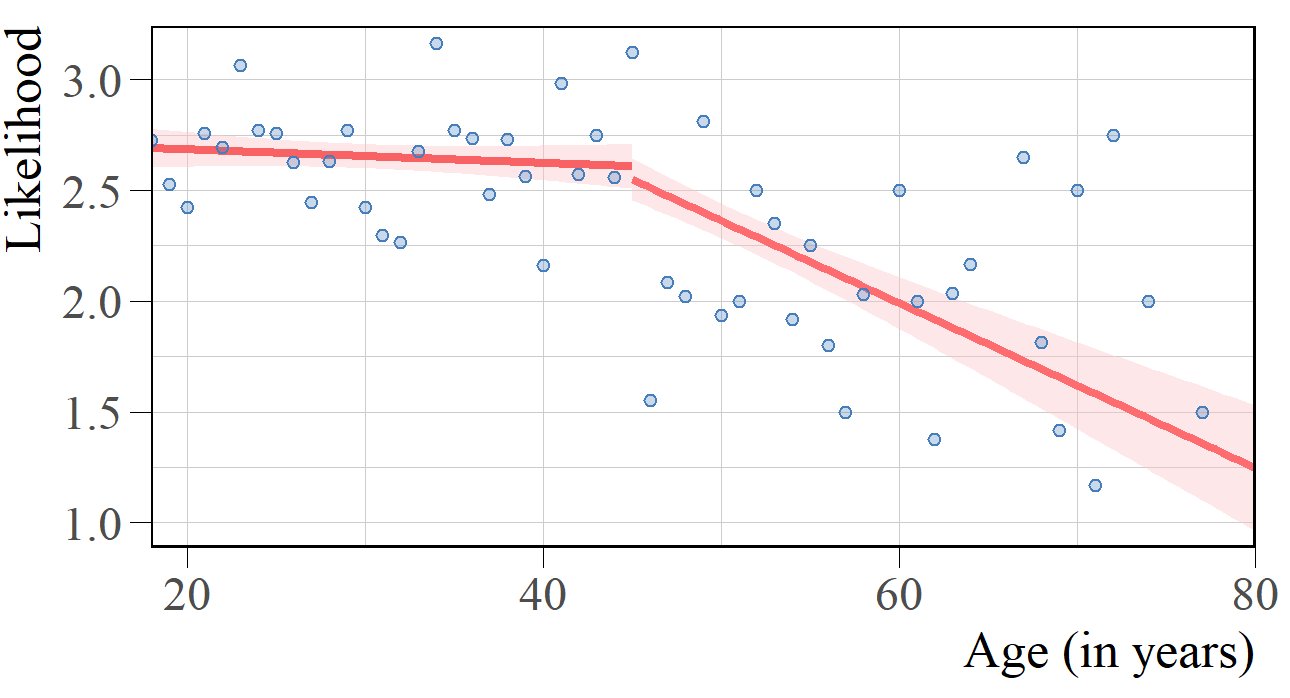}
\captionstyle{\raggedright}
\hangcaption
\caption{Piecewise relationship between consideration likelihood and age}
\label{fig:age}
\centering
\end{figure}

Preliminary inspections of the relationship between \textit{age} and the dependent variable reveal a non-linear relationship as shown in Figure \ref{fig:age}.
Accordingly, the covariate for \textit{age} is included in the model as a piecewise function, equal to zero prior to 45 years of age and increasing linearly after the age of 45. In other words, given a negative coefficient for this piecewise \textit{age} function, respondents over 45 are increasingly less likely to consider on-demand modes of travel for essential-purpose trips compared to younger individuals ($\beta$ = -0.047 , p = 0.000). In terms of interpretation, this implies that a respondent at the age of 67 is estimated to be a full consideration level lower than individuals under 45 years of age.

\subsubsection{Typical Modes of Travel}
The model also controls for respondents' self-reported typical modes of travel, which is likely to be a source of endogeneity in the experiment, especially for those who consider hailed ride services as a typical mode of travel. Indeed, the coefficient for habitual use of hailed ride services (compared to walking, cycling or transit) has the greatest magnitude among control covariates and is highly significant ($\beta$ = +1.073 , p = 0.000). On the other hand, the magnitude for private vehicles is negative ($\beta$ = -0.457 , p = 0.003).

\subsubsection{Tech-Savviness \& Attitudinal Constructs}
Similar to the effect of being a typical user of hailed ride services, it is hypothesized that having positive attitudes towards technology and being comfortable using technology would result in higher consideration of on-demand services overall, whether traditional or autonomous, and in higher consideration of new technologies, including autonomous cars. Accordingly, the latent construct \textit{tech-savviness} is estimated using exploratory and confirmatory factor analyses as shown in TABLE \ref{tab:latentvar}. The construct is significantly measured by the four indicator statements (p = 0.000) in TABLE \ref{tab:latentvar} with acceptable to good reliability according to the fit indices (RMSEA < 0.08, CFI > 0.95, TLI > 0.95) \cite{brown93, hu99, schumacker04}.

Estimated threshold parameters are omitted for brevity given that threshold parameters do not have a direct interpretation. Factor scores are calculated using the Empirical Bayes Modal method for each individual to be used \underline{serially} in the regression model.

\begin{table}[H]
\captionstyle{\raggedright}
\hangcaption
\caption{Confirmatory Factor Analysis Model for Latent Attitudes$^{\mbox{\ding{67}}}$}
\label{tab:latentvar}
\begin{tabular}{p{0.6\linewidth}ccc}
\Xhline{1.5pt}
\multicolumn{4}{|l|}{\textbf{\textit{Robust Model Statistics:}} \textit{Chi-squared test statistic: 86.099} \hspace{8 pt} \textit{Comparative Fit Index (CFI): 0.972}}                                                                     \\
\multicolumn{4}{|l|}{\hspace{120 pt} \textit{Degrees of Freedom: 19} \hspace{51 pt} \textit{Tucker-Lewis Index (TLI): 0.958}}                                                                                                                  \\
\multicolumn{4}{|l|}{\hspace{120 pt} \textit{Root Mean Square Error of Approximation (RMSEA): 0.072}}                                                                                                             \\ \Xhline{1.5pt}
\multicolumn{1}{|l|}{Indicator}                                                            & Estimate & z-value & \multicolumn{1}{c|}{p-value} \\ \Xhline{1.5pt}
\multicolumn{4}{|l|}{\textbf{Tech-Savviness}}                                                                                       \\ \hline
\multicolumn{1}{|>{\raggedright}p{0.63\linewidth}|}{\hangindent=25pt\hspace{10 pt}Technology is changing society for the better.}                                      & \hspace{2pt}0.740$^{***}$      & 27.28   & \multicolumn{1}{c|}{0.000}   \\
\multicolumn{1}{|>{\raggedright}p{0.63\linewidth}|}{\hangindent=25pt\hspace{10 pt}I am excited to learn about new technologies in the market.}                          & \hspace{2pt}0.916$^{***}$      & 43.81  & \multicolumn{1}{c|}{0.000}   \\
\multicolumn{1}{|>{\raggedright}p{0.63\linewidth}|}{\hangindent=25pt\hspace{10 pt}I pay more to get more technologically advanced products.}                          & \hspace{2pt}0.694$^{***}$      & 26.32  & \multicolumn{1}{c|}{0.000}  \\
\multicolumn{1}{|>{\raggedright}p{0.63\linewidth}|}{\hangindent=25pt\hspace{10 pt}I use the internet daily for chatting and entertainment.}                          & \hspace{2pt}0.502$^{***}$      & 12.29  & \multicolumn{1}{c|}{0.000}  \\\Xhline{1.5pt}
\multicolumn{4}{l}{\scriptsize \textit{$^{***}$ = significant at 0.01 level; $^{**}$ = significant at 0.05 level; $^{*}$ = significant at 0.10 level}}  \\
\multicolumn{4}{l}{\scriptsize \textit{\textsuperscript{$^{\mbox{\ding{67}}}$} Ordinal threshold effects are estimated but omitted from table for concision}}  
\end{tabular}
\end{table}

In line with the hypothesis, \textit{tech-savviness} has a positive and significant coefficient ($\beta$ = +0.413 , p = 0.000), indicating an increased likelihood of using on-demand services for essential trips for respondents who are considered to be tech-savvy.

The attitudinal construct for \textit{environmental consciousness} was also tested in the regression model but found to have an insignificant effect on the dependent variable ${LL}_{it}$.

\subsubsection{Threshold Parameters}
Similar to the threshold parameters in the confirmatory factor analysis in TABLE \ref{tab:latentvar}, the threshold parameters in TABLE \ref{tab:dndresults} are - for the most part - considered \textit{nuisance parameters} with limited value for interpretation but rather necessary for estimation \cite{greene10}. Nonetheless, the thresholds are all significant, indicating clear separation between likelihood levels in the dependent variable.

\section{Segmentation Analysis}
The latter section presented the difference-in-difference model results, showing that the COVID-19 pandemic had a significant effect on the consideration of autonomous vehicles. Nonetheless, these model results have an ambiguous interpretation of the effect of individual characteristics, such as mode choice or age, on the impact of COVID-19 on the future adoption of autonomous vehicles. In order to assess these effects in a more direct and intuitive manner, the model presented in the previous section is estimated for different segments of the sample.

\begin{figure}
\includegraphics[scale=.95]{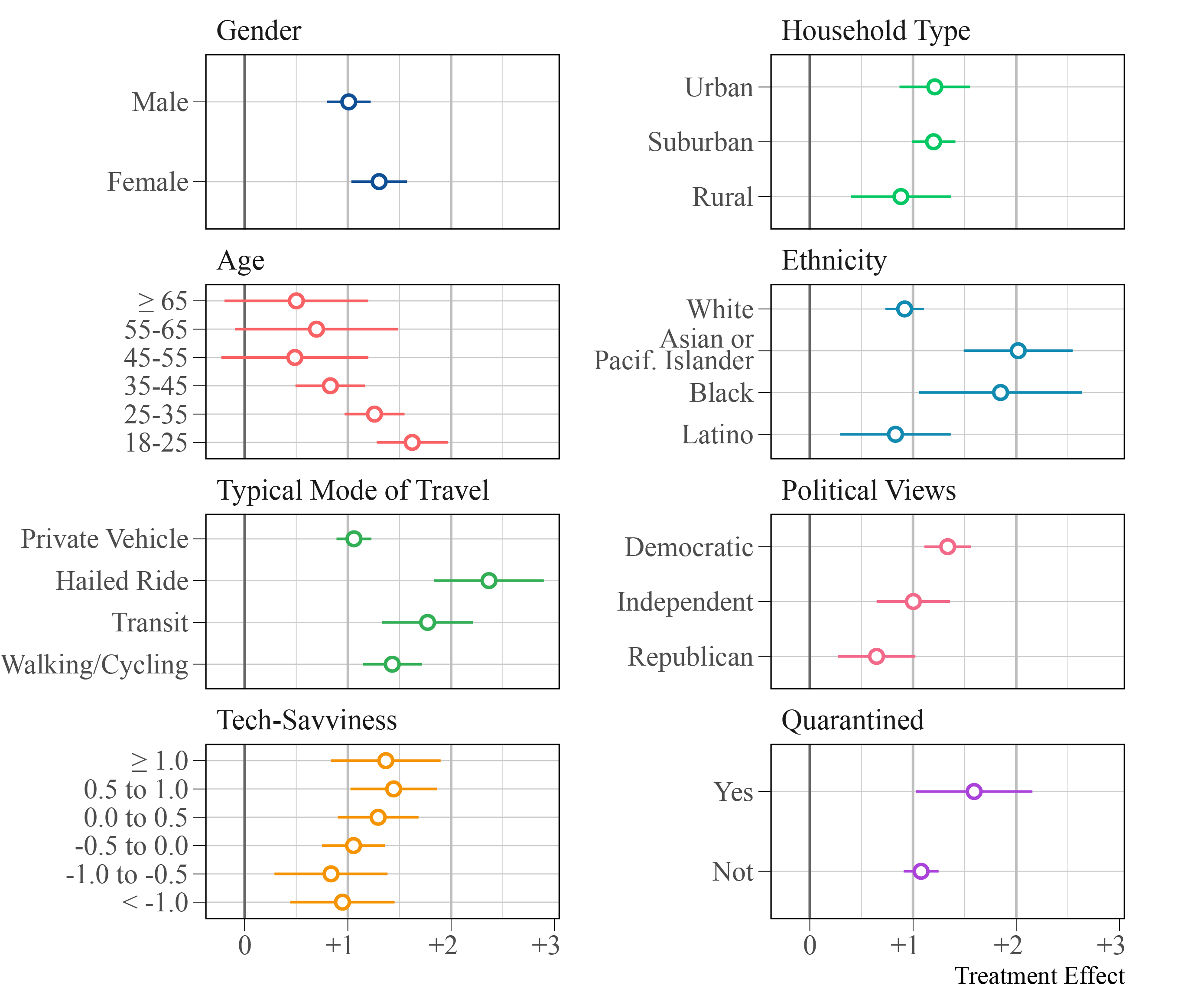}
\captionstyle{\raggedright}
\hangcaption
\caption{Treatment effects from the segmented models of individual characteristics on willingness to adopt autonomous vehicles as a result of COVID-19 (with 95\% confidence intervals)}
\label{fig:dndcovarresults}
\centering
\end{figure}

The treatment effects from the segmented models are summarized in Figure \ref{fig:dndcovarresults}, controlling for the same covariates presented in TABLE \ref{tab:dndresults}. Results are presented with the 95\% confidence intervals.

As defined in the previous sections, the \textit{treatment effect} in the y-axis of Figure \ref{fig:dndcovarresults} illustrates the change in likelihood of adopting AVs as a result of the COVID-19 pandemic. Here, the values 0 to +3 represent the change in this likelihood on a 5-point Likert scale (\textit{very unlikely}, \textit{unlikely}, \textit{neither likely nor unlikely}, \textit{likely}, \textit{very likely}). For example, a value of +1, as is the case for \textit{male} respondents, indicates that the pandemic has resulted in an average increase in likelihood of using AVs by 1 unit on the 5-point scale. In other words, all else constant, a male respondent who is originally \textit{neither likely nor unlikely} to use AV would shift to being \textit{likely} instead as a result of the pandemic. The figure also reveals - given the overlapping confidence intervals - that there is no significant gender difference in the pandemic-triggered consideration of AV use.

Looking at Figure \ref{fig:dndcovarresults}, the increase in consideration of AVs as a result of the COVID-19 pandemic is positive and significant for most segments, with the exception of respondents aged 45 or above. For some features there is little distinction, suggesting that regardless of differences in tech-savviness, gender or urban/rural household location, individuals are more likely to adopt AVs in the future. Instead, for a number of factors like age, typical travel mode and political views, there are structural differences within the category.

Looking at age, the treatment effect of COVID on AV consideration is declining with increasing age. While the magnitude of the treatment effect is 1.62 for respondents age 18 to 24, it declines to 0.83 for respondents age 35 to 44 and becomes insignificant for respondents older than 45. These results reveal that the increased propensity to adopt new technologies observed among younger individuals, a phenomenon which has been reported in several studies \cite{chen11}, remains true for AVs. This result is also consistent with pre-pandemic findings by Lavieri et al. \cite{lavieri17}.

Considering ethnicity and race, Asian and Pacific Islander respondents have a significantly greater increase in likelihood of using AVs owing to the pandemic compared to White or Latino respondents. Black respondents also experience a similar boost in AV consideration but have less prominent separation, with only 0.10 level of significance compared to White respondents. Ni\~no et al. \cite{nino21} found that ethnic minorities were more likely to be fearful of COVID-19 and to perceive it as a threat, which could explain why Asian, Pacific Islander and Black respondents are more likely to consider AVs as a safer mode of transport due to its no-contact advantage.

Considering the typical travel mode of respondents, results show that drivers would experience a significant increase in consideration of AVs due to the pandemic. That effect is however less than half that found among respondents who typically travel by hailed ride services. Transit and actives mode users (walking or cycling) fall in between these positions. This suggests that more shared transportation experiences lead to a higher shift in AV consideration. This trend is consistent with higher COVID-19 risk perception when using public modes of transport compared to private modes \cite{barbieri2021impact}.

With regard to political views, a declining treatment effect is observed as respondents lean Republican as opposed to leaning Democrat. This effect is in line with observations by Mack et al. \cite{mack21} who found that political moderates and liberals have higher AV adoption intentions than conservatives. The treatment effect could be explained by the fact that Republicans were less likely to abide by social distancing guidelines than individuals leaning Democrat \cite{hsiehchen20} and therefore feel less of a need for automated cars. Models for other political views, such as Libertarian or Liberal, are not estimated given the small number of observations for those segments in the sample.

The differences in treatment effects for other attributes, mainly gender, household location type, quarantine experience and tech-savviness, are insignificant. In the case of gender, some previous studies indicate that male gender is positively correlated with autonomous vehicle adoption  \cite{payre14,kyriakidis15,lavieri17} but this effect has also been observed to be insignificant \cite{payre14,kyriakidis15}. Tech-savviness displays a general positive correlation with AV consideration, but there is no significant difference in the magnitude of treatment effects across different levels of tech-saviness. This observation is at odds with the hypothesis that more tech-savvy individuals would have a greater shift in consideration of AVs due to the pandemic, but is consistent with the work of Koul and Eydgahi who studied the effect of technophobia (the reverse of tech-saviness) on autonomous vehicle adoption finding that it was insignificant in the model although there was a weak, negative correlation \cite{koul20}. Conversely, Lavieri et al. found that tech-saviness increased propensity to adopt AVs \cite{lavieri17}. 
Finally, individuals who have had to quarantine during the pandemic are noticeably more likely to consider AVs in the future on average, yet this effect is also insignificant, even at a significance level of 0.10.

\section{Conclusion}
Given the push of the COVID-19 pandemic towards the use of digitization and numerous technologies in order to ameliorate the impacts of restricted mobility and social distancing, this study analyzes the impact of COVID-19 on consideration and future adoption of new technologies. Specifically, this paper assesses the change in consideration and perception of autonomous vehicles due to the pandemic.

Using data from a U.S. sample of 691 respondents, this study finds that, indeed, the COVID-19 pandemic has a significant positive effect on the intention to use autonomous vehicles in the future. Specifically, younger, left-leaning and frequent users of shared modes of travel would become more likely to use autonomous vehicles once offered. Other variables have a more nuanced impact. Ethnicity is associated with different degrees of shifting intentions, with Black, Asian and Pacific Islander respondents having a more pronounced positive shift compared to White and Latino respondents. Contrary to what is commonly hypothesized, the extent to which one is tech-savvy has a limited effect on the shift in consideration due to the pandemic. Similarly, women and men experience a similar pandemic-induced increase in consideration.

Understanding the effects of these attributes on the changes in consideration of AVs is important for policy making, as these effects provide a guide to predicting adoption of autonomous vehicles - once available - and to identify segments of the population likely to be more resistant to adopting AVs. Identifying these segments is important especially given the potential benefits of AVs to marginalized segments of the population, such as seniors who are hypothesized to benefit from the increase accessibility AVs would provide them.

\section{Limitations \& Future Work}
This study is not without limitations. Given that autonomous vehicles are not yet available for adoption, respondents may face some difficulty in anchoring their potential experience with the mode. Nonetheless, this study is designed to minimize this issue. Specifically, respondents are presented with questions about a shared autonomous vehicle service, which is not dissimilar to ridehailing or taxi services, albeit without a driver. By reducing the difference to solely the presence of a driver, this study aims to minimize any anchoring issues of the pandemic on this consideration process. Additionally, this study uses convenience online sampling recruitment, resulting in a sample that is more representative of left-leaning, lower-income and younger respondents. These sociodemographic features are, however, controlled for in the models presented in this study, reducing potential biases.

Another limitation relates to the use of cross-sectional data to measure change in consideration over time. Ideally, this study would have used longitudinal data collected at two points in time, before the pandemic and during. This data could have informed whether varying levels of COVID-related infections or hospitalizations relate to different levels of openness towards autonomous vehicle use. Nonetheless, given the unpredictable nature of the pandemic, acquiring data in such a longitudinal manner is challenging. While this study acknowledges the biases inherent in asking respondents to recall their past consideration of using a mode the data still offers valuable insight into the impact of the pandemic on this consideration process.

Further research is needed to more fully understand the role of the pandemic in shaping future adoption of transportation innovations like automated driving. 
A first important avenue of future research is to better account for the dynamic nature of the decision-making surrounding the use of new mobility options in the COVID-19 era. At least three areas of behavior require careful data-collection, experiment design and modeling: (1) daily travel decisions have seen major shifts with increased use of (private) cars and active modes as well as decreased use of transit; (2) risk-perceptions and protective behaviors are shaped by vaccine penetration, restrictive policies and case-rates which are all undergoing constant change; (3) the view of AV driving technology itself is shaped by personal and media-diffused experiences, technological maturation and changing business models (such as sharing vs. ownership). Future research should aim to better capture the dynamics both within and across these areas. Future AV adoption will contend with general daily travel trends (durability of pro-car attitudes), the ongoing impact of pandemic risk behavior (objective changes and subjective risks) and the state of AV development (acceleration in development and deployment fostering new demand).

Second, our results suggest a general positive pandemic-era shift due to the latent construct measuring tech-savviness. As the pandemic persists and morphs, we invite further research on how AV adoption is shaped by COVID-19 in the long run. Promising areas include identifying latent factors related to future work-from-home and remote office policies that are likely to alter home-location and daily travel patterns. Attitudes and aspirations around future work should be tested in mobility innovation adoption models. A second promising area is to identify latent factors related to pandemic risk. In particular, we suspect that AV adoption might be driven both by enthusiasm (closely related to our tech-savviness concept) and fear (avoiding viral exposure). Further work is needed to better capture the mix of push and pull factors driving adoption of AVs in addition to views of different use models (shared fleet versus private ownership).

Third, this work highlights a new aspect to investigate further, namely the impact of political beliefs on willingness to adopt new technologies. Although this work finds that respondents who lean Democrat have a greater consideration of autonomous vehicles due to the pandemic, further research is needed to know whether this observation is due to COVID-specific political attitudes (e.g. regard for social distancing guidelines) or whether this effect will persist beyond the presence of COVID mitigation measures. This question is particularly important as increasing political polarization may lead to a divide between segments of the population according to willingness to adopt new technological developments, including autonomous vehicles.

\bigskip

\section{Acknowledgements}
\hspace{\parindent}This research was partially supported by the US National Science Foundation Career Award No. 1847537, the Northwestern Buffett Institute Global Impacts Graduate Fellowship and the National Science Foundation Graduate Research Fellowship Program DGE-184216. The study was approved by Northwestern’s Institutional Review Board under study number STU00212452.

\bigskip

\section{Author Contributions}
The authors confirm contribution to this paper as follows: study conception and design: MS, AS; data collection: MS, AS; analysis and interpretation: MS, EZ, AS; draft manuscript preparation: MS, EZ, AS. All authors reviewed the results and approved the final version of the manuscript.
\newpage

\bibliographystyle{trb}
\bibliography{AV_COVID}
\end{document}